\documentclass[12pt]{article}

\pdfoutput=1
\usepackage{definitions}
\usepackage{heppennames2}
\usepackage{Note}

\title{Muon Beam at the Fermilab Test Beam Area}

\date{\today}
\def\MyNoteNumber{FERMILAB-TM-2627-E}

\addauthor{Dmitri Denisov}{\institute{3}}
\addauthor{Valery Evdokimov}{\institute{2}}
\addauthor{Strahinja Lukić \thanks{slukic@vinca.rs}}{\institute{1}}
\addauthor{Predrag Ujić}{\institute{1}}

\addinstitute{1}{Vinča Institute, University of Belgrade, Serbia}
\addinstitute{2}{Institute for High Energy Physics, Protvino, Russia}
\addinstitute{3}{Fermilab, Batavia IL, USA}

\abstract{The intensities and profiles of the muon beam behind the beam dump of the Fermilab test beam area when the facility is running in the ``pion'' beam mode are measured and summarized in this note. This muon beam with momenta in the range $10\div50\unit{GeV/c}$ provides an opportunity to perform various measurements in parallel with other users of the test beam area.}

\graphicspath{{./figures/} }

\begin{document}

\titlepage

\section{Introduction}
\label{sec:intro}

The Fermilab Test Beam Facility (FTBF) is a high-energy test beam facility at Fermilab devoted to Detector R\&D \cite{ftbf}. The facility uses two versatile beamlines (MTest and MCenter) to produce a variety of particle types with wide range of energies for testing particle detectors. The primary beam consists of 120~GeV protons with intensities of $1 - 300\unit{kHz}$. This beam can also be used to create secondary beams of pions, muons and/or electrons with momenta between 1 and 60~GeV/c. 

The beam dump of the MTest beam area is built from concrete slabs with a total thickness of 3.2~m. When secondary beams are used in the MTest area, muons generated in pion decays penetrate behind the beam dump. This area thus has the potential to be used for tests when a relatively pure broadband muon beam is required, with no interference with other experiments in the MTest area.

The setup used for the measurement of the muon beam intensity and profile is described in Sec.\ \ref{sec:setup}, beam conditions are described in Sec.\ \ref{sec:beam}, results are given in Sec.\ \ref{sec:results} and conclusions are given in Sec.\ \ref{sec:conclusions}.

\section{Measurement setup}
\label{sec:setup}

\begin{figure}
\centering
   \includegraphics[width=140mm]{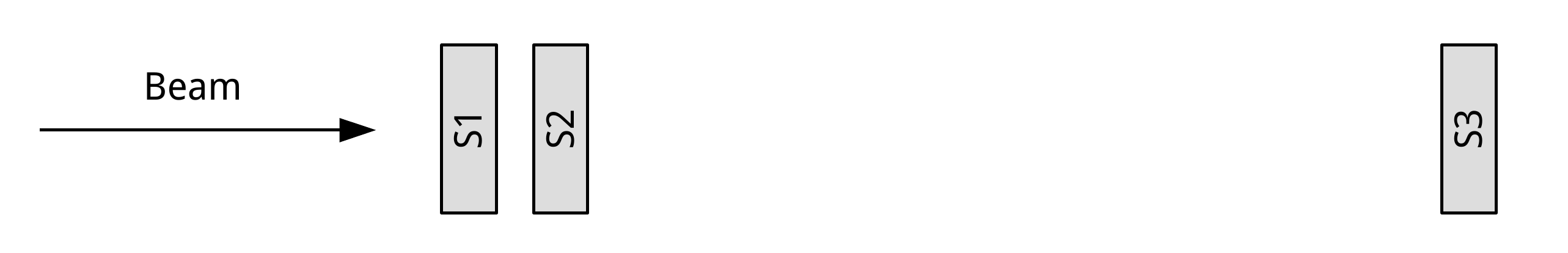}
   \caption{\label{fig:telescope} The scinitllator telescope setup used 
              for the measurement of the muon beam profile and intensity.}
\end{figure}

Plastic scintillation counters, denoted \sone, \stwo and \sthr, aligned along the beam direction were used for the measurement of the muon rate during the accelerator spills (Fig.\ \ref{fig:telescope}). The counters were mounted on an aluminum frame to allow displacing them in the horizontal and vertical directions perpendicular to the beam. This way it was possible to scan the beam profile in several points in both directions. Two counters were used in the position \sone for different runs:

\begin{enumerate}
  \item Counter \sonea has an area of $16\times24\unit{cm}^2$ and was mounted 
    with the longer side in the vertical direction,
  \item Counter \soneb has an area of $10\times15\unit{cm}^2$ and was mounted 
    with the longer side in the vertical direction. 
\end{enumerate}
  
Counter \stwo has an area of $2.7\times40\unit{cm}^2$ and was mounted always with the shorter side in the direction of the scan, i.e.\ vertically for the hirozontal scan and horizontally for the vertical scan. Counter \sthr has an area of $16\times24\unit{cm}^2$, and was mounted with the longer side in the vertical direction. All counters have a thickness of 1.2~cm. The signals from the counters were processed by a constant threshold discriminator to produce logical timing pulses. A logical coincidence unit was used to register signal coincidence between the counters within the coincidence time of 80~ns. 


The threshold on all discriminators was set to 30~mV and the high voltage at the counters was set to ensure most probable muon signal amplitude of about 100~mV.

Test beam is delivered to the MTest area in spills with a duration of 4.2~s. The spills are repeated every 60 seconds. In the measurements presented here, the scaler registering the muon count was always reset less than 5~s before the incoming spill, and the number of counts was registered 10~s after the scaler reset. To estimate the background counting rate, background counts were registered during 50~s between the spills and then averaged for each counters configuration separately. 

The four used configurations are summarized in Table \ref{tab:coinc}. The background counting rates are very low and are neglected in the following.

\begin{table}
  \caption{\label{tab:coinc}Coincidence requirements used in the measurements and the corresponding effective detection areas and the measured background rates.}
  \centering
  \begin{tabular}{ c | c | c | c | c }
    \hline
    Configuration & Trigger &     Area      & Background  & Comment \\
         \#       &         & $\unit{cm}^2$ & $\unit{s}^{-1}$ &    \\
    \hline
    1 & $\sonea \wedge \sthr$ & 384 & 0.02  & \\
    2 & $\soneb \wedge \sthr$ & 150 & 0.007 & \\
    3 & $\stwo \wedge \sthr$  & 43  & 0.005 & $y$-scan, \stwo mounted horizontally \\
    4 & $\soneb \wedge \stwo \wedge \sthr$ & 40 & 0.008 & $x$-scan, \stwo mounted vertically \\
	\end{tabular}
\end{table}

\section{Beam conditions}
\label{sec:beam}

\begin{figure*}
\centering
   \includegraphics[width=.96\textheight, angle=90]{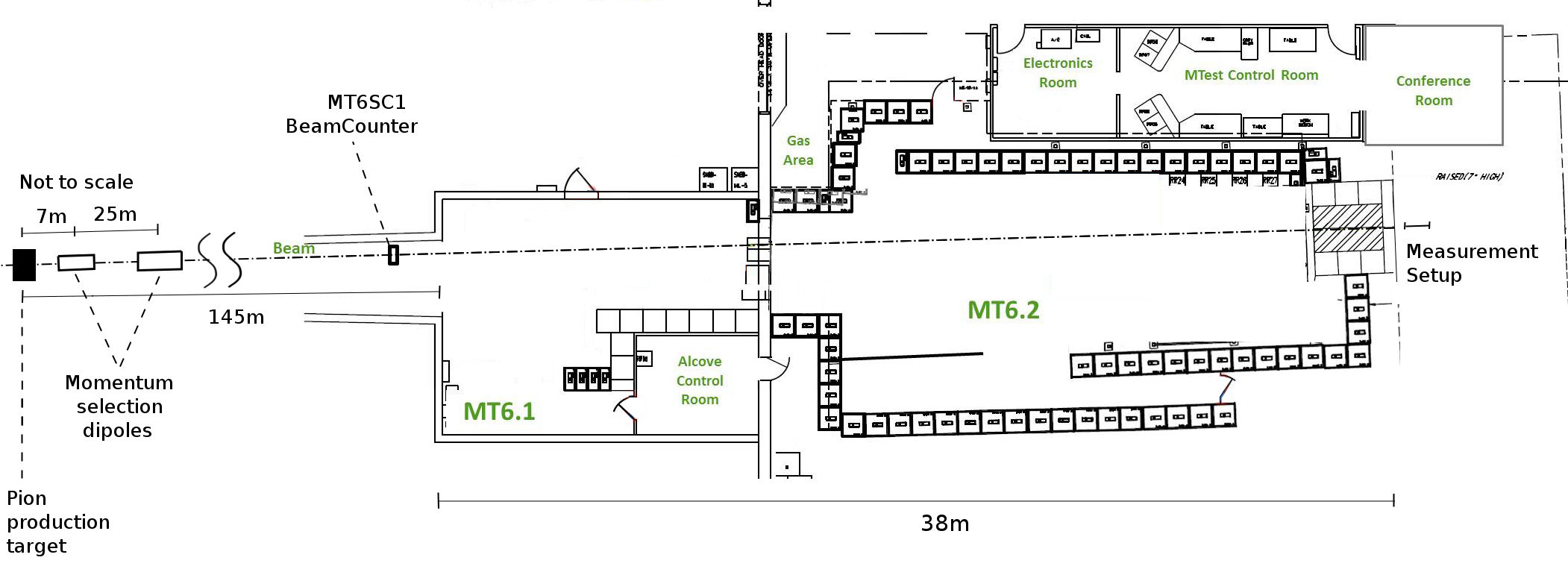}
   \caption{\label{fig:map} Map of the MT6 test beam area with beam path distances. Location of the MT6SC1 beam counter is also shown (see text).}
\end{figure*}


The measurements presented here have been performed with the test beam in the "low-energy pion" mode and pion beam momentum up to 28~GeV/c. For the low-energy pion mode the secondary-beam production target is located 145~m upstream from the entrance to the test-beam area MT6. The dipoles for the momentum selection are located 7~m and 25~m downstream from the target. The length of the MT6 area including the beam dump is 38~m. Beam absorbers in the test beam area were moved out of the beam so that the pion beam continues to the beam dump. The muon measurement setup was located immediately behind the beam dump. Map of the test beam area with a sketch of the beam path distances is shown in Fig.\ \ref{fig:map}. 

The total length from the production target to the muon measurement setup is 183~m. Decay path length of a 16~GeV/c pion is 890~m, and that of a 28~GeV/c pion is 1560~m. Thus, depending on the pion beam energy, only between 1.5 and 3\% of pions decay before the momentum selection and most muons arriving to the beam dump are produced from a narrow-band pion beam after momentum selection in the dipoles. The muon energy spectrum is thus governed by the decay kinematics of the quazi-monoenergetic pion beam and extends from approximately one half of the pion beam energy to the full pion beam energy.

During the measurements, the sPHENIX calorimeter was tested in the MT6.2 area in front of the beam dump. The electromagnetic calorimeter of sPHENIX consists of tungsten absorbers and plastic scintillation counters for a total of 18 radiation lengths (\xo). The hadronic calorimeter of sPHENIX is based on steel absorbers and plastic scintillation counters for a total of 35 \xo. The total number of nuclear interaction lengths of the sPHENIX calorimeter is 6 \cite{phenix15}. The thickness of the beam dump is 30 \xo of shielding concrete. The energy loss for minimum-ionizing particles across the sPHENIX calorimeter and the beam dump is 2.2~GeV. Thus the minimum momentum for a muon to reach the measurement setup is close to 3~GeV/c.

\subsection{Muon beam intensity}
\label{sec:beam:I}

Beam intensity in the test beam area is routinely monitored using several scintillation counters at various positions along the beam line. The counter MT6SC1 is located at the entrance to the MT6 area (Fig.\ \ref{fig:map}). The sensitive area of MT6SC1 is $10\unit{cm}^2$ and its thickness is 6~mm. Pion beam intensities with up to 1 million counts per spill in the MT6SC1 counter were achieved in the MT6 beam area. Most measurements presented here were performed with beam intensities of $\lesssim 10^4$ counts per spill in MT6SC1. For consistence of beam shape measurements, all measured muon beam intensities were multiplied by a factor $10\times10^3/N_\text{MT6SC1}$, where $N_\text{MT6SC1}$ is the average number of counts per spill in the MT6SC1 counter during the measurement in question. All presented muon beam intensities and fluxes have been normalized in this way.


\subsection{Muon beam size}
\label{sec:beam:size}

\subsubsection{Pion decay kinematics}

The last beam collimator is located around 60~m downstream from the secondary-beam production target. Thus for the majority of pion decays, muons emitted under the full $4\pi$ solid angle in the decaying pion rest frame can reach the beam dump. Since the muon momentum in the pion rest frame is 30~MeV/c, the muon angular distribution extends from 0 to around 1~mrad for the 28~GeV/c pion beam. The longitudinal location of the pion decays is exponentially distributed between the production target and the beam dump with the mean decay path length of $\gamma c\tau$, where $\tau$ is the pion decay time in the rest frame. The resulting radial distribution of muon tracks has a cusp-like shape in the detection plane if only the pion decay kinematics are considered. Monte-Carlo (MC) simulation of the kinematics for a 28~GeV/c \PGp beam leads to the FWHM of the cusp of $x_\text{RMS,decay} = 13\unit{cm}$, if the collimators are neglected. In the case of 16 and 24~GeV/c pion beams, the FWHM of the cusp is 25 and 17~cm, respectively. 

\subsubsection{Multiple scattering and pion beam spot size}

Considering the total material thickness of $83\unit{\xo}$ in the path of the muon beam, the contribution of the multiple scattering to the muon beam size was estimated by MC using the prescription of the Particle Data Group \cite{pdg15} based on the formula by Highland \cite{Highland} for the dependence of the scattering angle on the muon momentum. The result is $\text{FWHM}_{x,\text{MS}} = 3\unit{cm}$ for the 28~GeV/c pion beam and 5~cm for the 16~GeV/c pion beam. 

The pion beam size during the measurements was about 3~cm at the MT6SC1 counter.

\section{Results}
\label{sec:results}

The center of the muon beam was found at a height of 156~cm from the floor and a horizontal distance of 130~cm from the protruding shielding wall at the right side of the area behind the beam dump facing the beam. The muon beam was scanned using counters of progressively smaller sizes with increasing pion beam energy.

\begin{figure}
\centering
   \includegraphics[width=0.6\textwidth]{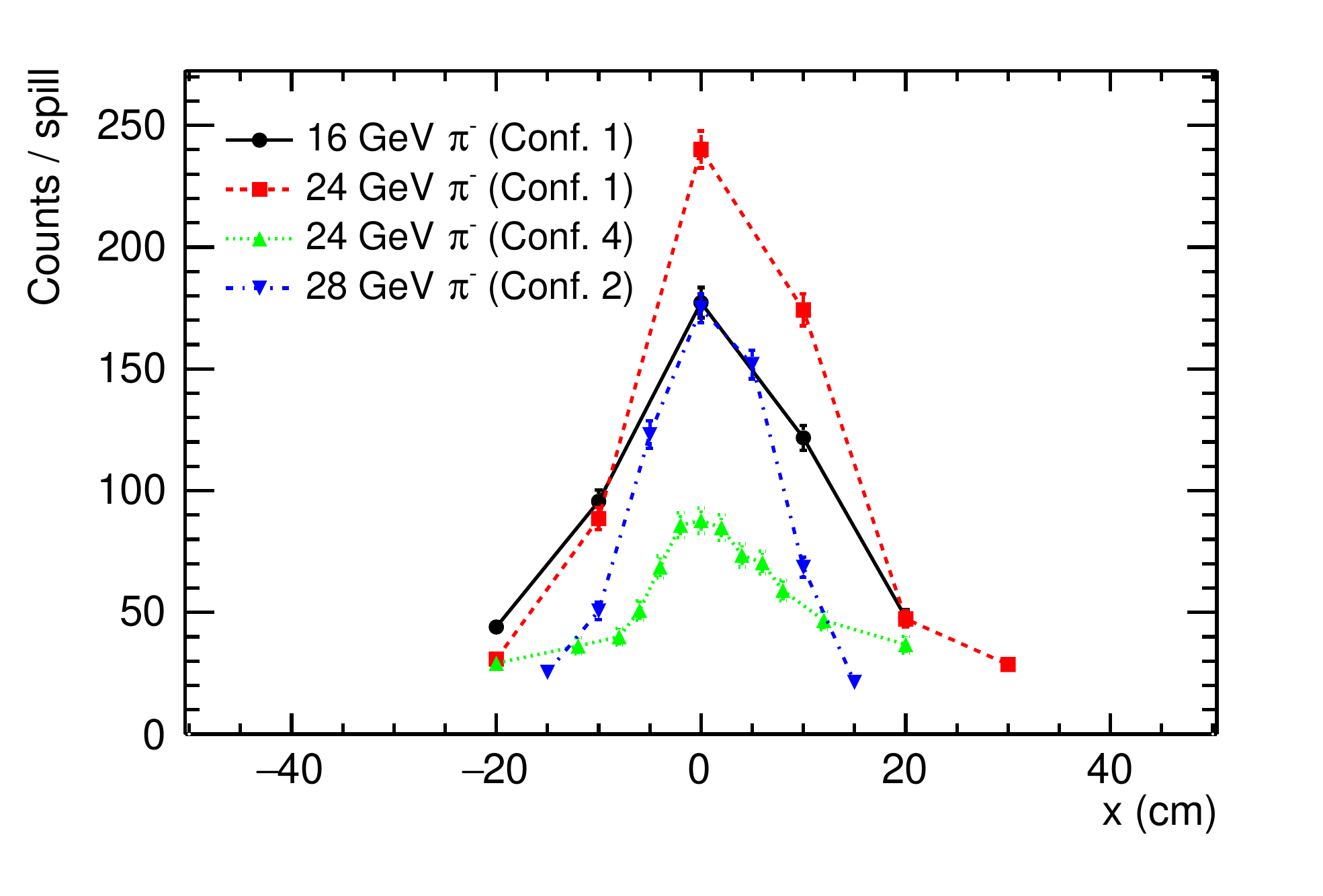}
   \caption{\label{fig:xscan} Muon beam profile scan in the horizontal direction for several pion beam momenta and various configurations of the counters described in Table \ref{tab:coinc}.}
\end{figure}

\subsection{Beam profile}

Results of the muon beam profile scan in the horizontal ($x$) direction are shown in Fig.\ \ref{fig:xscan}. The scan was performed in the low-energy pion mode with pion beam momenta of 16, 24 and 28~GeV/c. A peak is clearly visible at each energy. Scans made with the pion beam momentum of 24~GeV/c, covering a wider area, reveal also a broad asymmetric distribution underneath the peak. This distribution is presumably formed by muons generated upstream from the momentum-selection dipoles. The ratio of the peak to the broad distribution varies and may depend on the beam tuning, collimator openings, etc.

\begin{figure}
\centering
   \includegraphics[width=0.6\textwidth]{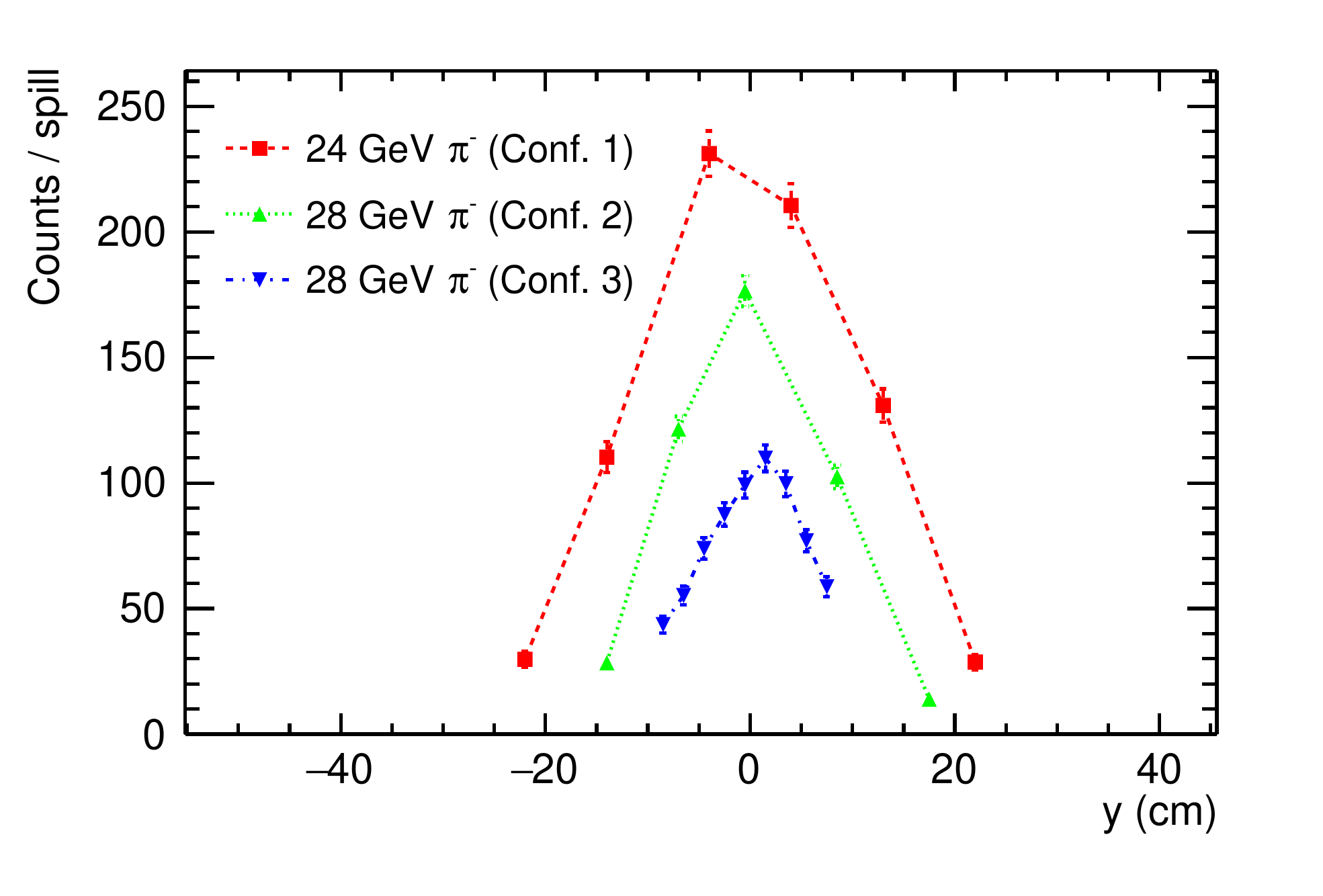}
   \caption{\label{fig:yscan} Muon beam profile scan in the vertical direction for several different pion beam energies and various configurations of the counters described in Table \ref{tab:coinc}. The reference point for the $y$-coordinate is the beam center, located at 156~cm from the floor.}
\end{figure}

Results of the muon beam scan in the vertical ($y$) direction are shown in Fig.\ \ref{fig:yscan}. The scan was performed in the low-energy pion mode with pion beam momenta of 16, 24 and 28~GeV/c. 

\begin{figure}
\centering
   \includegraphics[width=0.6\textwidth]{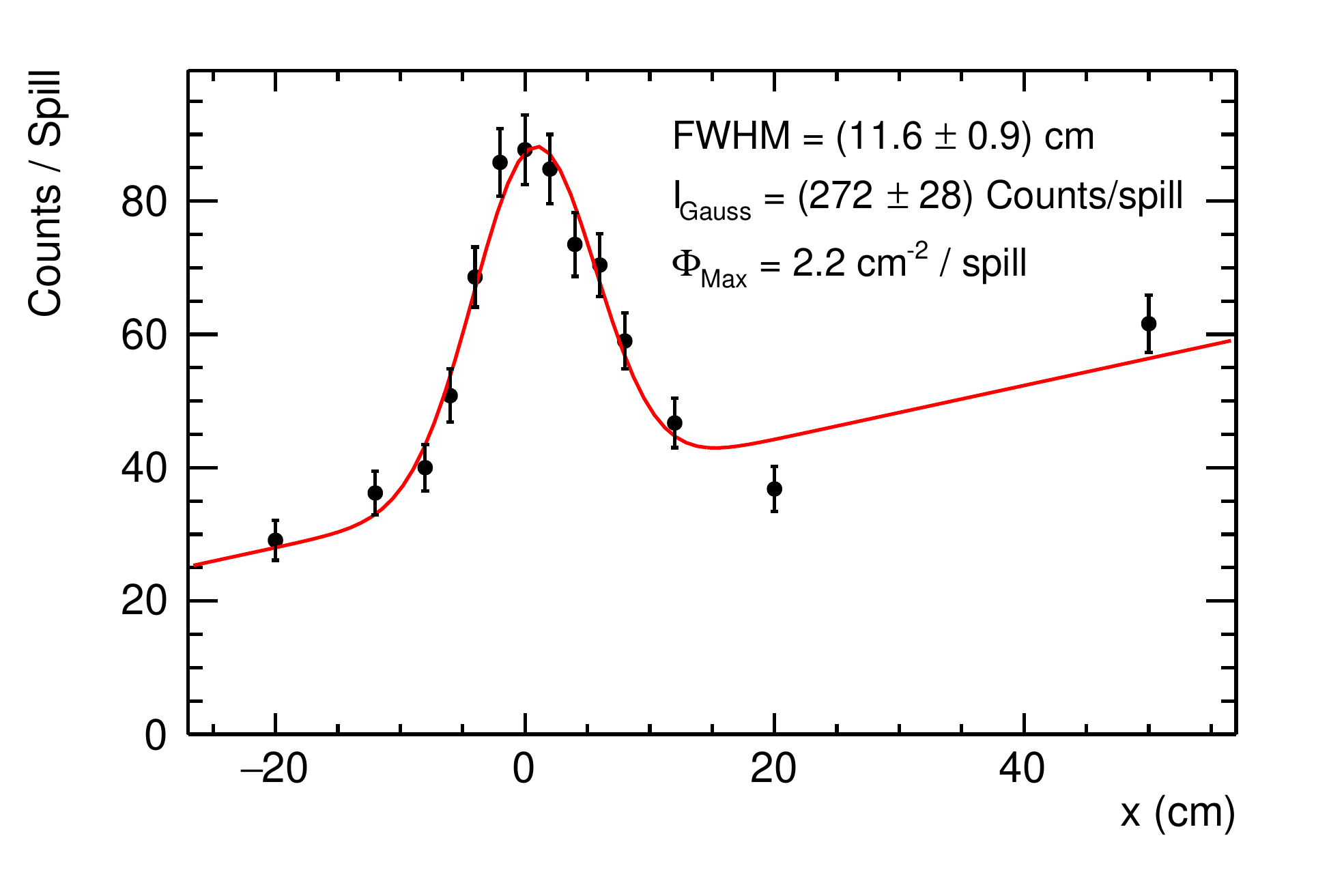}
   \caption{\label{fig:fit} Scan in the horizontal direction of the muon beam profile for the 24~GeV/c pion beam. A fit with the Gaussian superimposed with a linear function is also shown. The beam intensity in the Gaussian part of the distribution is extracted from the fitted parameters, taking into account the width of the counter and correcting for the finite coverage in the $y$-direction.}
\end{figure}

Muon beam profile scan in the horizontal direction for the 24~GeV/c \PGpm beam, fitted with a Gaussian superimposed with a linear function is shown in Fig.\ \ref{fig:fit}. 

\subsection{Beam angle}

\begin{figure}
\centering
   \includegraphics[width=140mm]{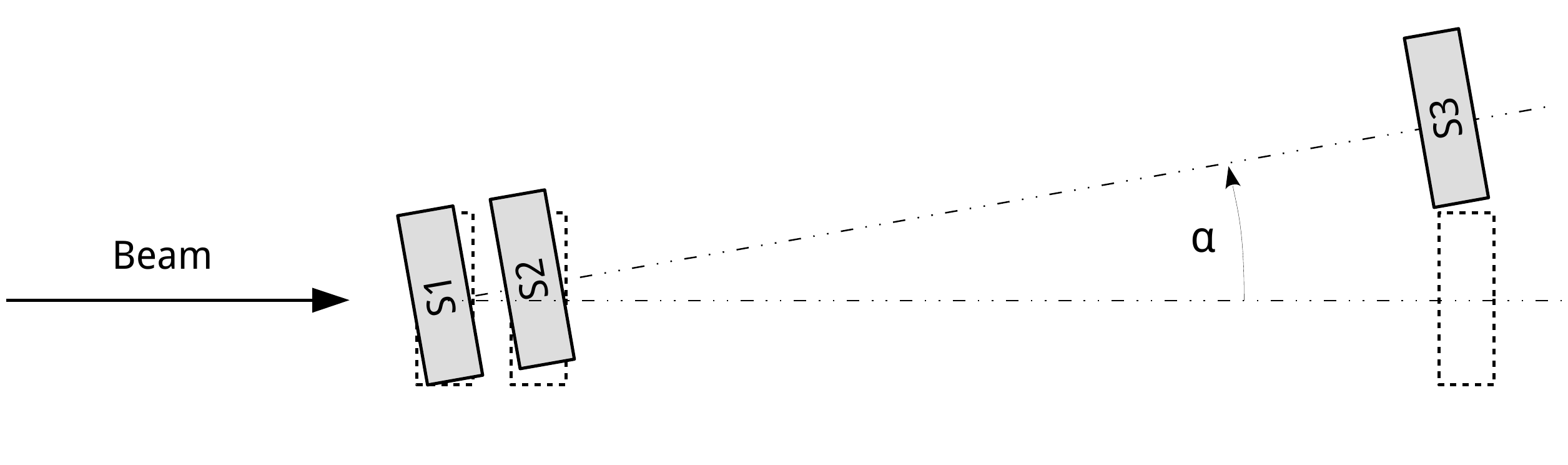}
   \caption{\label{fig:asketch} Sketch of the angular scan.}
\end{figure}

\begin{figure}
\centering
   \includegraphics[width=0.6\textwidth]{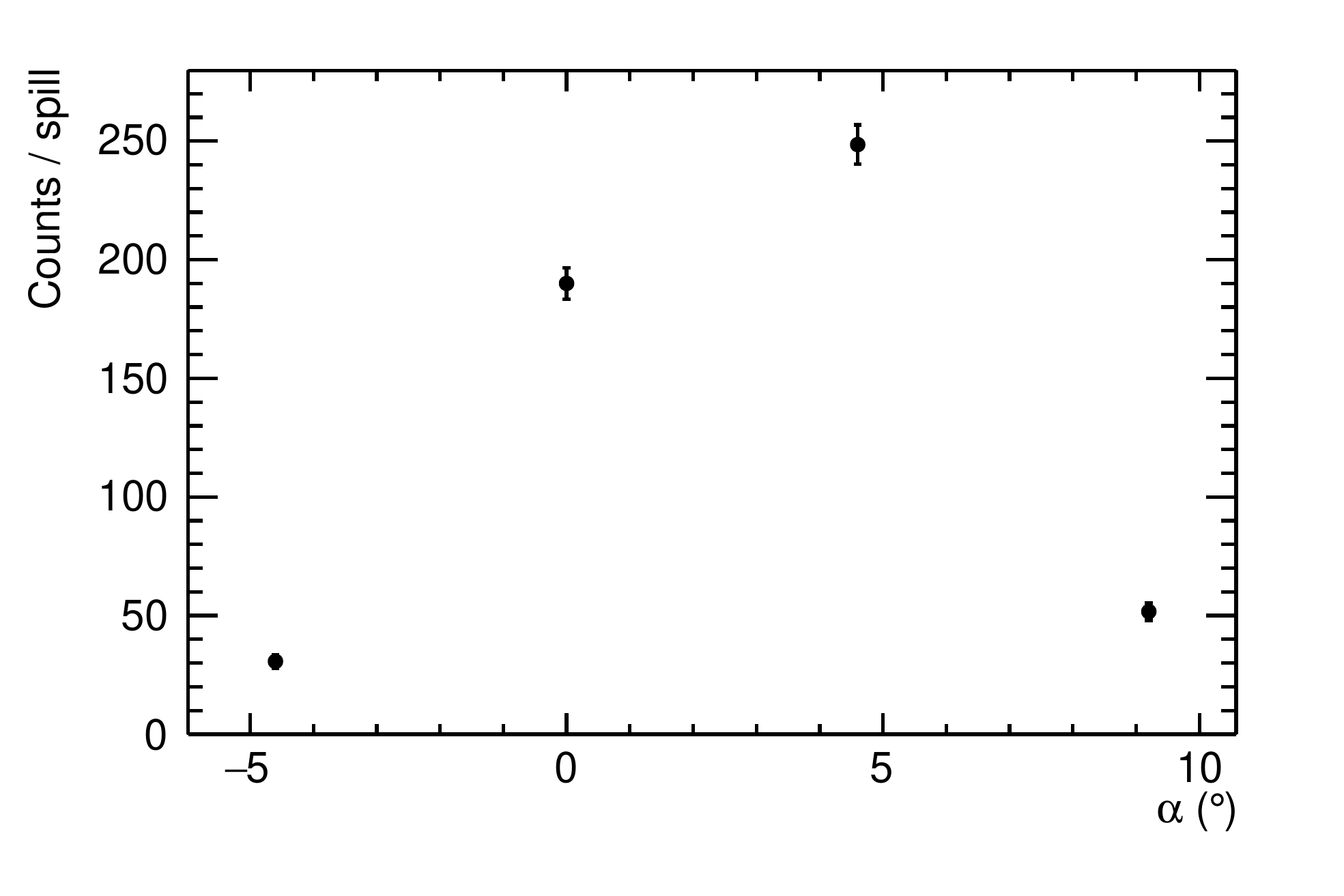}
   \caption{\label{fig:ascan} Scan of the angular distribution of the muons for the \PGpm beam momentum of 28~GeV/c.}
\end{figure}

A scan of the angular distribution of the muons for the 28~GeV/c \PGpm beam was performed by rotating the setup around the base of the counter \sone for an angle $\alpha$ between $-4.6\degrees$ and $9.2\degrees$ relative to the axis perpendicular to the rear wall of the beam dump. Positive angles correspond to counterclockwise rotations seen from above (Fig.\ \ref{fig:asketch}). Results are presented in Fig.\ \ref{fig:ascan}. The average muon beam angle weighted by the number of counts per spill is $(2.6\pm0.2)\degrees$.

\subsection{Beam intensity}

The effective detection area in configurations 1 and 2 covers large part of the beam cross-section area in which the muon flux is significantly lower than at the profile maximum. The maximum muon fluxes are measured with configurations 3 and 4 and are given in Table \ref{tab:fluxes}. Fluxes for 24 and 28~GeV/c pion beam momenta have been measured during the $x$ and $y$-scans, and the flux for the 16~GeV/c pion beam momentum has been measured at a single point. All fluxes have been normalized to correspond to the pion beam intensity producing $10\times10^3$ counts per spill in the MT6SC1 counter. 

\begin{figure}
\centering
   \includegraphics[width=0.6\textwidth]{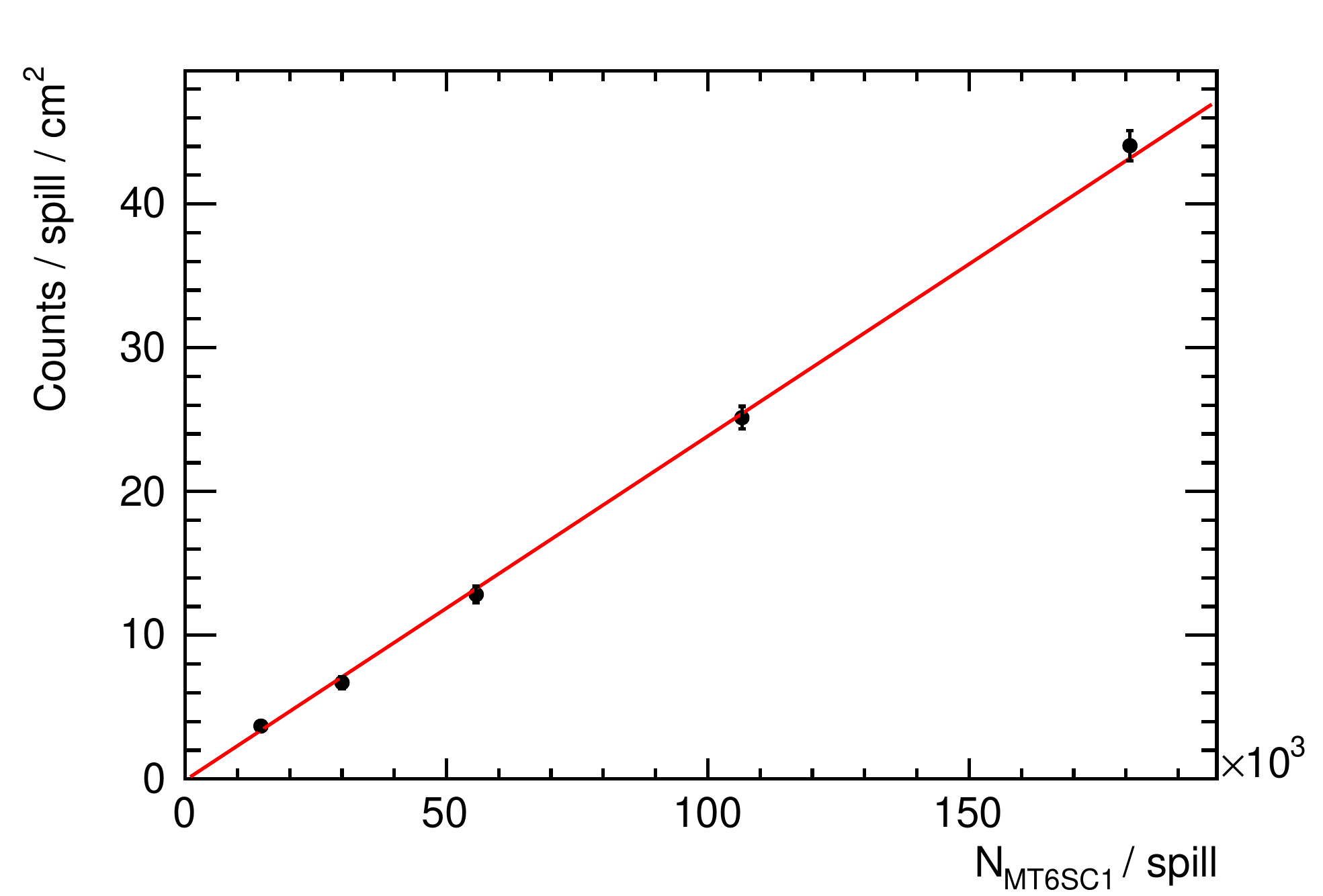}
   \caption{\label{fig:ramp} Dependence of the muon count in the measurement setup vs.\ pion beam intensity as measured in the MT6SC1 test beam counter for secondary beam momentum of 28~GeV/c.}
\end{figure}

\begin{table}
  \caption{\label{tab:fluxes}Muon beam intensities behind the beam dump. Flux data refer to the measurements at the beam center for the given pion beam energy averaged over the effective area of the counter configuration used ($\sim 40\unit{cm}^2$). Intensities have been normalized to correspond to the pion beam intensity producing $10\times10^3$ counts per spill in the MT6SC1 test beam counter.}
  \centering
  \begin{tabular}{ c | c }
    \hline
    $p_\pi$ & $\Phi_{\PGm,\text{max}}$ \\
       GeV/c  & $\unit{cm}^{-2}/\unit{spill}$ \\
    \hline
    16  &  1.1  \\
    24  &  2.2  \\
    28  &  2.6  \\
	\end{tabular}
\end{table}

Dependence of the muon beam intensity vs.\ pion beam intensity, as measured MT6SC1 test beam counter, was measured for secondary beam momentum of 28~GeV/c. Results, expressed as the number of muons per spill per $\text{cm}^2$, are shown in Fig.\ \ref{fig:ramp}. Straight line fit to the data is also shown.

\section{Conclusions}
\label{sec:conclusions}

Muon beam profile and intensity have been measured behind the beam dump of the test beam area at FTBF. Beam spot with FWHM between 10 and 20~cm was observed for different pion beam energies. Muon fluxes from 1 to 3 counts per spill per $\text{cm}^{2}$ have been measured with pion beam intensity corresponding to $10\times10^3$ counts per spill in the MT6SC1 counter. The muon beam intensity increases linearly with the pion beam intensity at least up to $180\times10^3$ counts per spill in the MT6SC1 counter. Muon flux up to $44\unit{spill}^{-1}\unit{cm}^{-2}$ was measured in these conditions.

This muon beam has excellent properties to perform studies of various particle detectors, especially for the detection of muons.

\printbibliography[title=References]

\end{document}